%% file: main.tex
\documentclass[10pt,conference]{IEEEtran}

\usepackage{cite}
\usepackage{graphicx}
\usepackage{amsmath,amssymb}
\usepackage{booktabs}
\usepackage{multirow}
\usepackage{makecell}
\usepackage{xcolor}
\usepackage{colortbl}
\usepackage{adjustbox}
\usepackage[most]{tcolorbox}
\usepackage{url}
\usepackage[hidelinks]{hyperref}

\newtcolorbox{answerbox}{colback=black!4,colframe=black!55,boxrule=0.5pt,arc=1pt,
  left=3pt,right=3pt,top=1.5pt,bottom=1.5pt,before skip=5pt,after skip=5pt,fontupper=\small}
\newtcolorbox{insightbox}{enhanced,colback=black!6,colframe=black!50,boxrule=0pt,leftrule=1.6pt,
  arc=0pt,left=3pt,right=3pt,top=1.5pt,bottom=1.5pt,before skip=5pt,after skip=5pt,fontupper=\small}
\newcommand{\op}[1]{($\rightarrow$\,\hyperref[sec:open]{OP#1})}

\newcommand{\vv}{\mathbf{v}}
\newcommand{\vh}{\hat{\mathbf{v}}}
\newcommand{\Acc}{\mathit{Acc}}

\begin{document}

\title{Narratives Mislead: An Empirical Study of How LLMs Misjudge CVSS Severity in Bug Bounty Reports}
% \author{\IEEEauthorblockN{Anonymous Author(s)}}
\author{
\renewcommand{\arraystretch}{1.2}
\begin{tabular*}{\textwidth}{@{\extracolsep{\fill}}cccc@{}}
Ruitao Feng & Yuekang Li & Yao Zhang \\
 Southern Cross University &    University of New South Wales &  Tianjin University \\
 ruitao.feng@scu.edu.au &  yuekang.li@unsw.edu.au & zzyy@tju.edu.cn 
 \end{tabular*}
}

\maketitle

\input{sections/0-abstract}
\input{sections/1-introduction}
\input{sections/2-background}
\input{sections/3-study-design}
\input{sections/4-rq1}
\input{sections/5-rq2}
\input{sections/6-rq3}
\input{sections/7-structured-reports}
\input{sections/7-discussion}
\input{sections/8-related-work}
\input{sections/8b-open-problems}
\input{sections/9-conclusion}
\input{sections/10-data-availability}
\input{sections/11-ack}

\bibliographystyle{IEEEtran}
\bibliography{references}

\end{document}

%% file: sections/0-abstract.tex
\begin{abstract}
Bug bounty platforms receive vulnerability reports faster than vendors can triage them, and Large Language Models (LLMs) are increasingly considered for assigning the Common Vulnerability Scoring System (CVSS) severity that drives prioritization. Prior evaluations score short CVE descriptions and report accuracy, but deployment depends on \emph{how} models err, not only how often. CVSS gives each metric a formal meaning that departs from how researchers describe impact in narrative reports, and model rationales suggest that errors concentrate where the two diverge; we ask whether this pattern is systematic. From 4,281 publicly disclosed reports mined from Huntr and HackerOne, we build a benchmark of 240 Huntr reports with vendor-validated CVSS vectors and evaluate eight LLMs under four prompting strategies and fine-tuning, measuring errors by severity class and score and analysing sub-vectors on individual reports. The best prompting configuration reaches only 49.58\% level accuracy, ten points above always predicting the majority level, and fine-tuning a 7B model raises this to at most 56.25\%, an optimistic upper bound. Errors are directional: models overestimate far more often than they underestimate, and they almost never recognize low-severity vulnerabilities. We trace these errors to three semantic boundaries that the specification draws and narrative reports blur: potential versus realized impact, impact magnitude versus authorization scope, and uncertainty versus the worst case. Because the violations originate partly in the report, we close with an exploratory case study: on four reports, a boundary-aligned, exploit-free report structure lowered a frontier model's score where the narrative invited an unclaimed account takeover, but left the metrics a report does not settle open, and the two published labels disagreed on three of the four reports.
\end{abstract}


%% file: sections/1-introduction.tex
\section{Introduction}
\label{sec:intro}

% P1 Background: triage is the bottleneck, and its input is sensitive
Vulnerability triage on bug bounty platforms has become a throughput problem. Bug Bounty Programs (BBPs) reward external researchers for reporting vulnerabilities~\cite{maillart2017given}, and platforms such as HackerOne~\cite{hackerone} and Huntr~\cite{huntr} route large volumes of reports to vendors~\cite{piao2024study,akgul2023bug}. Each report must be reproduced and assigned a severity, typically a Common Vulnerability Scoring System (CVSS) score~\cite{FIRST2023}, before it can be prioritized~\cite{zhang2015survey}. This step is slow and inconsistent: report quality varies widely~\cite{bettenburg2008makes,al2018friendly}, scoring needs security expertise, and manual assessment does not scale~\cite{han2017learning}. Automating it is attractive~\cite{walkowski2021automatic}, but the input is unusually sensitive: an undisclosed report describes an unpatched, often exploitable flaw, so vendors are reluctant to send it to third-party services and may only be able to use models they run themselves.

% P2 Milestones
Early approaches learned to predict CVSS scores from the short descriptions in vulnerability databases, using shallow CNNs~\cite{han2017learning}, multi-target learning~\cite{spanos2018multi}, character-level CNNs~\cite{nakagawa2019character}, and distilled transformers~\cite{kai2023vuldistilbert}. LLMs promise to remove task-specific training~\cite{yin2024multitask}, and two recent studies evaluate them on this task. Marchiori et al.~\cite{marchiori2025can} found that even the best model falls below 50\% accuracy on Privileges Required and Availability when deriving CVSS~v3.1 vectors from CVE descriptions. Jafarikhah et al.~\cite{jafarikhah2025description} evaluated six LLM families on over 31,000 CVEs, reported about 79\% average per-metric accuracy for GPT-5, and attributed many remaining errors to descriptions lacking context, recommending richer inputs such as proof-of-concept exploits.

% P3 Problem: two mismatches
These studies leave two questions open that matter most in a real triage pipeline. The first is an \emph{input} mismatch: they score CVE descriptions of a few hundred characters, whereas a triage assistant reads the raw report, a long researcher-written narrative with reproduction steps, a proof of concept, and the author's own account of impact. Such reports are the ``richer context'' prior work calls for, but richer is not neutral: their authors' rewards typically grow with severity~\cite{hackerone_bountytables}, and impact sections routinely describe what a flaw \emph{could} lead to. The second is an \emph{error-structure} mismatch: prior work reports per-metric accuracy and, at most, a preference for majority classes~\cite{jafarikhah2025description}, but not the \emph{direction} of errors or their relation to how CVSS defines each metric. For deployment this is decisive: symmetric noise degrades triage gracefully, whereas systematic inflation silently reorders the remediation queue.

\begin{figure*}[t]
  \centering
  \includegraphics[width=0.8\textwidth]{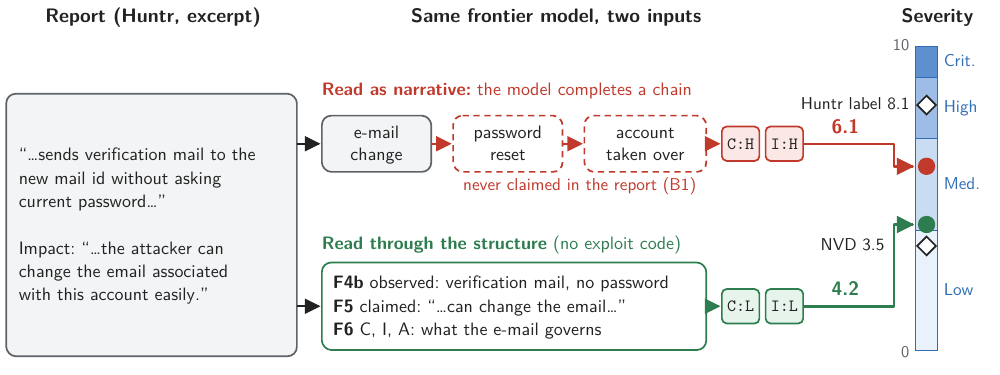}
  \caption{One report, two readings. The report shows that an e-mail change needs no password. Read as narrative, a frontier model chains it to an account takeover the reporter never claims and scores 6.1 (C:H/I:H); read through a structure that separates the reporter's claims from the observed effect, the same model scores 4.2 (C:L/I:L), closer to the NVD label (Sect.~\ref{sec:running}).}
  \label{fig:motivation}
\end{figure*}

% P4 Insight
Model rationales suggest that the structure of these errors is not accidental, because CVSS is not a free-form judgment. Its metrics carry formal definitions that depart from everyday meaning: Scope records whether an exploit crosses an \emph{authorization boundary}, not how extensive the damage is; the impact metrics capture what the report \emph{substantiates}, not every downstream consequence; and missing information is unknown, not the worst case~\cite{FIRST2023}. Bug bounty reports, in contrast, use a narrative language in which ``could lead to remote code execution'' conveys urgency. An LLM reasoning from the narrative rather than the specification would then fail predictably (Fig.~\ref{fig:motivation}): errors should concentrate on metrics whose formal and everyday meanings diverge, point upward, and trace back to identifiable report features. If so, the gap lies partly in the report, and when only small local models may see the report, its \emph{representation} is a lever worth exploring.

% P5 This paper
We examine this question in an empirical study: on a benchmark of 240 Huntr reports whose complete CVSS vectors were agreed during vendor--researcher triage, we evaluate eight proprietary and open-source LLMs, including the 7--8B models a vendor can host itself, under four prompting strategies and fine-tuning, against TF-IDF and majority-class baselines, and measure errors by severity class and by continuous score and its direction, and, for reports analysed in depth, by sub-vector. We ask how accurate predictions are and in which direction they err (\textbf{RQ1}), where errors concentrate (\textbf{RQ2}), and which semantic boundaries are violated and why (\textbf{RQ3}). As an exploratory extension, we try a report structure aligned with the violated boundaries on four representative reports, with a capability-limited and a frontier model.

% P6 Headline results
LLM errors are structured rather than random. The best prompting configuration assigns the correct severity level to only 49.58\% of reports, ten points above always predicting Medium, and the best fine-tuned model 56.25\%. Every prompting configuration overestimates at least 45\% of reports, up to 42\% of predictions miss by two or more severity levels, and, under few-shot prompting, no model recognizes more than one of the 25 Low reports. We trace these errors to three semantic boundary violations (Fig.~\ref{fig:boundaries}): potential impact read as realized (B1), impact magnitude read as a scope change (B2), and missing information read as the worst case (B3). Capability does not dissolve them: in our running examples a frontier tool-using model also inferred an account takeover that the report never claims (Fig.~\ref{fig:motivation}). In the exploratory case study, a boundary-aligned, exploit-free report structure dampened that error (6.1 to 4.2) and made the metrics a report leaves open explicit, but did not settle them.

% P7 Contributions
\smallskip\noindent\textbf{Contributions.}
\begin{itemize}
\item \textbf{A diagnosis of how LLMs misjudge CVSS on real reports.} Errors are directional and specification-grounded; we characterize them as three semantic boundary violations, using benchmark-wide statistics and sub-vector analysis of individual reports.
\item \textbf{A sub-vector-level benchmark and evaluation.} A dataset of 4,281 publicly disclosed bug bounty reports, including a benchmark of 240 with vendor-validated CVSS vectors, released with the paper, and a three-level error framework that records whether a prediction is wrong, in which direction, and, where a rationale is available, at which sub-vector.
\item \textbf{An exploratory case study on report structure.} A first attempt at an exploit-free report structure aligned with the three boundaries, tried on four reports, which shows where restructuring helps and where it does not, and five open problems.
\end{itemize}

%% file: sections/2-background.tex
\section{Background}
\label{sec:background}

\paragraph{Vulnerability reports.} A bug bounty report documents a vulnerability found by an external researcher. It typically contains a description, reproduction steps or a proof of concept (PoC), and an impact statement written by the researcher~\cite{malladi2019bug,farris2018vulcon}. Vendors use the report to reproduce the issue and to assign a severity; on Huntr, the agreed CVSS vector is published with the disclosed report. Supplement~A walks through how each sub-vector is derived from one report.

\begin{figure*}[t]
  \centering
  \includegraphics[width=0.8\textwidth]{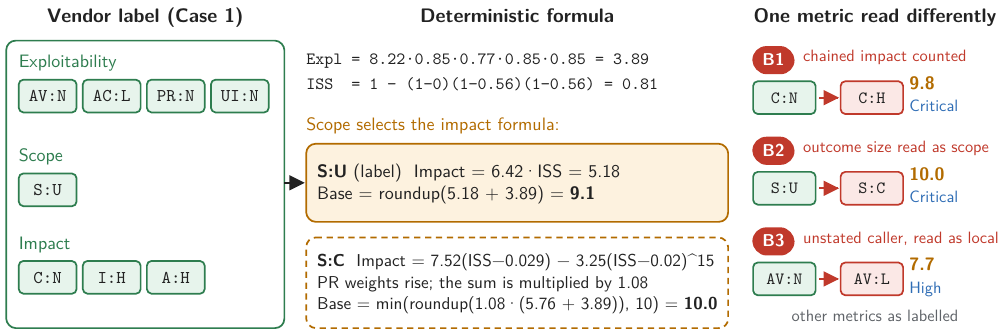}
  \caption{The CVSS computation on a real label (Case~1 of Sect.~\ref{sec:rq3}). The vector determines the score through a fixed formula in which Scope selects the impact equation. Right: reading a single metric differently moves the score by up to 1.4 points and can change the severity level. B1 and B2 push the score up; for B3, the report does not state who calls the vulnerable function, so AV can be read either way, and the label and some models resolved it in opposite directions, which is why the structure of Sect.~\ref{sec:structure} records it as unknown.}
  \label{fig:cvss}
\end{figure*}

\paragraph{CVSS.} The CVSS Base Score, maintained by FIRST~\cite{FIRST2023}, rates the intrinsic severity of a vulnerability from 0 to 10. It is computed from eight \emph{sub-vectors} (Fig.~\ref{fig:cvss}, left): four exploitability metrics, namely Attack Vector (AV), Attack Complexity (AC), Privileges Required (PR), and User Interaction (UI); Scope (S); and three impact metrics, namely Confidentiality (C), Integrity (I), and Availability (A). Each metric takes a qualitative value (e.g., C:H) that maps to a fixed weight, and the weights enter a deterministic formula. Scope switches the formula: when an exploit can affect resources beyond the \emph{security authority} of the vulnerable component (S:C), the impact curve becomes steeper, PR weights increase, and the total is multiplied by 1.08. The score finally maps to a severity level: None (0.0), Low (0.1--3.9), Medium (4.0--6.9), High (7.0--8.9), or Critical (9.0--10.0). Two properties matter for this paper. First, every metric has a normative definition in the specification that an assessor must apply, not an everyday reading. Second, the computation from vector to score is fully deterministic, so any error in the final score originates either in the vector or in arithmetic, and one misread metric can move the score across a level boundary (Fig.~\ref{fig:cvss}, right).

%% file: sections/3-study-design.tex
\section{Study Design}
\label{sec:design}

\subsection{Task and Error Definitions}
\label{sec:task}

Given a report $r$, a predictor outputs a CVSS score $\hat{s}$; under chain-of-thought (CoT) prompting it also outputs a vector $\vh=(\hat{v}_{\mathrm{AV}},\dots,\hat{v}_{\mathrm{A}})$. The ground truth is the vector $\vv$ agreed during triage, with score $s=f(\vv)$, where $f$ is the CVSS~v3 formula, and level $L(s)$. We measure errors at three levels:
\begin{itemize}
\item \textbf{Level.} The prediction is correct if $L(\hat{s})=L(s)$. The signed level distance $\Delta_L$ is the difference of the two levels' ranks (None\,=\,0, \dots, Critical\,=\,4); $|\Delta_L|\geq2$ is a high-risk misclassification.
\item \textbf{Score.} The signed error $e=\hat{s}-s$; $e>0$ is overestimation.
\item \textbf{Sub-vector.} For each metric $m$ we order its values by their contribution to severity, e.g., $\mathrm{N}\prec\mathrm{L}\prec\mathrm{H}$ for C/I/A, $\mathrm{U}\prec\mathrm{C}$ for S, and $\mathrm{P}\prec\mathrm{L}\prec\mathrm{A}\prec\mathrm{N}$ for AV. A metric is \emph{over-assessed} if $\hat{v}_m\succ v_m$ and \emph{under-assessed} if $\hat{v}_m\prec v_m$.
\end{itemize}
Because $f$ is deterministic, we can also recompute $\tilde{s}=f(\vh)$ from a model's own vector. If $\tilde{s}\neq\hat{s}$, the model reasoned to a vector and then miscomputed it; this separates \emph{assessment} errors from \emph{arithmetic} errors.

\subsection{Data}
\label{sec:data}

We collected publicly disclosed reports from HackerOne Hacktivity (disclosed between December 2016 and March 2024) and Huntr (from its launch in 2020 to March 2024), using a rate-limited crawler restricted to public pages. For each report we kept the full text (description, PoC, impact), the CVSS base score, the CWE identifier, and, for Huntr only, the complete CVSS vector. We removed reports under 100 words~\cite{zimmermann2010makes}; two researchers then independently excluded non-reports and reports whose technical content lived only in unavailable links, with a third resolving disagreements. Labels were never edited. Labels use CVSS~v3.0 or v3.1, whose Base equations differ only in the precise definition of rounding~\cite{FIRST2023}.

The \emph{benchmark} consists of 240 Huntr reports covering 59 CWE types, sampled to preserve the original distribution of severity levels and major CWE categories. A disjoint \emph{fine-tuning set} contains 821 further Huntr reports and 3,220 HackerOne reports, for 4,281 reports in total, covering 201 CWE types; only the Huntr reports carry full vectors. The benchmark's levels are Critical (43), High (77), Medium (94), Low (25), and None (1). At $n=240$ the standard error of an accuracy near 50\% is 0.031 (Supplement~C), so a 95\% interval for a single configuration spans about $\pm6$ percentage points; because all configurations score the same reports, this marginal interval is a conservative yardstick for comparing them. Ground-truth labels are the vectors agreed between vendor and researcher during triage; we discuss their subjectivity in Sect.~\ref{sec:threats}. We use only public pages, do not redistribute exploit code, and will release report identifiers, labels, prompts, and model outputs with the paper.

\subsection{Predictors}
\label{sec:predictors}

\paragraph{LLMs.} We evaluate eight models, covering proprietary and open-source families and several parameter scales: Deep\-Seek-R1~\cite{guo2025deepseek}, GPT-4o~\cite{hurst2024gpt}, GPT-3.5-turbo~\cite{openai_chatgpt}, LLaMA-3.1 in its 405B and 8B instruction-tuned variants~\cite{touvron2023llama}, and the instruction-tuned Qwen2.5-7B~\cite{qwen2.5}, Gemma-7B~\cite{team2024gemma}, and Mistral-7B~\cite{mistral2023}.

\paragraph{Prompting strategies.} Prompts are assembled from six modules: core instruction, output format, CoT steps, an annotated example, the input report, and a reasoning check (Supplement~B). \emph{0-shot} asks a ``computer security expert'' for a CVSS score in a fixed output format. \emph{0-shot CoT} adds three reasoning steps modelled on expert practice~\cite{wei2022chain,zhou2022least}: extract the vulnerability type, impact, and vector; assess each sub-vector; compute the score. \emph{Few-shot} adds one annotated example report, and \emph{few-shot CoT} combines both. DeepSeek-R1 already reasons internally, so it is evaluated without CoT modules.

\paragraph{Fine-tuning.} We fine-tune the four 7--8B open-source models on the fine-tuning set with LLaMA-Factory~\cite{zheng2024llamafactory}, using a 0-shot prompt, since CoT mainly benefits models above 100B parameters~\cite{wei2022chain} and HackerOne reports lack sub-vector labels. LLaMA-3.1-405B exceeded our hardware budget. Configurations were selected on the benchmark (Supplement~E), so fine-tuned results are an optimistic upper bound.

\paragraph{Baselines.} A \emph{majority-class} baseline always predicts Medium, the most frequent level in the benchmark (94 of 240 reports, 39.17\%). A \emph{TF-IDF + Random Forest} baseline, common in prior severity prediction work~\cite{kudjo2019improving,yosifova2021predicting}, uses unigram and bigram TF-IDF features (10,000 dimensions) and is trained on the fine-tuning set both as a regressor of the score and as a five-class classifier of the level.

\subsection{Metrics and Output Handling}
\label{sec:metrics}

At the level of severity classes we report $\Acc$ and per-class precision, recall, and F1. At the score level we report MAE, RMSE, over- and under-estimation rates, the rates of crossing one or two severity levels, and the distribution of absolute errors (definitions in Supplement~B). Scores are extracted with a regular expression from the required output format. Outputs that contain no score, refuse to answer, or address a different report are \emph{invalid}; $\Acc_{\mathrm{all}}$ counts them as wrong and $\Acc_{\mathrm{valid}}$ excludes them. Small models occasionally emit values outside $[0,10]$, which inflate MAE, RMSE, and the mean signed error; we therefore compare configurations by rates and error bands, which such outliers barely affect, and quote MAE and RMSE only for the best configurations. We refer to the 30 prompting configurations (eight models $\times$ four strategies, minus the two CoT strategies for DeepSeek-R1) and the four fine-tuned models together as the 34 configurations. Sub-vector-level analysis requires the reasoning that accompanies a vector, so we apply it to the reports analysed in depth (Sect.~\ref{sec:rq3}) and to the running examples (Sect.~\ref{sec:structure}).

%% file: sections/4-rq1.tex
\section{RQ1: How Accurate, and in Which Direction?}
\label{sec:rq1}

\paragraph{Accuracy is low and barely moved by prompting.} Table~\ref{tab:main} reports level accuracy. No configuration exceeds 49.58\% (GPT-4o with 0-shot CoT; LLaMA-3.1-405B with few-shot CoT), only 10.4 points above always predicting Medium (39.17\%) and 6.7 points above TF-IDF regression; the latter gap is barely larger than the $\pm6$-point uncertainty of the benchmark. For large models, prompting hardly matters: GPT-4o ranges from 44.58\% to 49.58\%, and adding examples \emph{lowers} its accuracy by about four points. Small models gain substantially from examples (Mistral-7B rises from 13.33\% to 36.25\%, Gemma-7B from 21.25\% to 42.08\%), but mostly because examples suppress invalid outputs: at 0-shot, LLaMA-3.1-8B refuses 70 of 240 reports and Mistral-7B answers a different question 116 times (Supplement~D). Examples also introduce a failure of their own: small models occasionally score the example instead of the target report.

\begin{insightbox}
\textbf{Insight 1: the ceiling is not in the prompt.} For large models the prompt moves accuracy by about five points, and the best result is ten points above always answering Medium. Once elicitation stops mattering, the residual error is set by what the report states, how its label was reached, or the model itself, none of which a prompt changes \op{1}.
\end{insightbox}

\begin{table}[t]
  \centering
  \caption{Severity-level accuracy (\%), $\Acc_{\mathrm{all}}$ / $\Acc_{\mathrm{valid}}$. Best LLM result per strategy in bold. ``--'': CoT modules not applied to the reasoning model.}
  \label{tab:main}
  \footnotesize
  \setlength{\tabcolsep}{3.4pt}
  \begin{adjustbox}{max width=\columnwidth}
  \begin{tabular}{@{}lcccc@{}}
    \toprule
    Model & 0-shot & 0-shot CoT & Few-shot & Few-shot CoT \\
    \midrule
    DeepSeek-R1      & 42.92/42.92 & -- & 45.42/45.42 & -- \\
    GPT-4o           & \textbf{48.75}/\textbf{48.75} & \textbf{49.58}/\textbf{49.58} & 44.58/44.58 & 45.83/45.83 \\
    GPT-3.5-turbo    & 45.00/45.00 & 42.50/42.50 & 35.00/38.01 & 28.75/39.66 \\
    LLaMA-3.1-405B   & 43.75/44.49 & 45.83/46.03 & \textbf{48.33}/\textbf{48.54} & \textbf{49.58}/\textbf{49.58} \\
    LLaMA-3.1-8B     & 19.17/34.59 & 30.83/42.05 & 38.75/39.74 & 37.08/38.20 \\
    Qwen2.5-7B       & 37.08/39.56 & 30.00/31.30 & 42.08/42.62 & 43.75/44.30 \\
    Gemma-7B         & 21.25/23.08 & 31.67/35.00 & 42.08/44.89 & 40.83/41.88 \\
    Mistral-7B       & 13.33/27.83 & 25.42/33.15 & 36.25/38.50 & 37.08/38.70 \\
    \midrule
    Majority class (Medium) & \multicolumn{4}{c}{39.17} \\
    TF-IDF+RF (reg./cls.) & \multicolumn{4}{c}{42.92 / 36.25} \\
    \bottomrule
  \end{tabular}
  \end{adjustbox}
\end{table}

\paragraph{Errors are large and directional.} Table~\ref{tab:direction} summarizes the score-level errors over the 30 prompting configurations. Even the best configuration deviates from the vendor score by 1.58 CVSS points on average (MAE), and RMSE exceeds MAE throughout, indicating a heavy tail: at least 14.6\% of predictions in every configuration miss by more than three points. The errors also have a direction. In every prompting configuration at least 45.4\% of reports are overestimated, and in the most inflationary one 91.0\% are, whereas no configuration underestimates more than 47.9\%. Between 5.0\% and 41.6\% of predictions miss the true level by two or more steps, e.g., a Low vulnerability scored High, the kind of error that silently reorders a triage queue.

\begin{table}[t]
  \centering
  \caption{Score-level errors: best and worst value over the 30 prompting configurations. MAE and RMSE are given for the best configuration only, because out-of-range outputs of small models inflate their maxima.}
  \label{tab:direction}
  \footnotesize
  \setlength{\tabcolsep}{5pt}
  \begin{adjustbox}{max width=\columnwidth}
  \begin{tabular}{@{}lcc@{\hspace{14pt}}lcc@{}}
    \toprule
    Metric & best & worst & Absolute error & min & max \\
    \midrule
    MAE (points) & 1.58 & -- & $\leq 0.5$ & 10.4\% & 26.7\% \\
    RMSE (points) & 2.06 & -- & 0.5--1.0 & 11.1\% & 25.5\% \\
    Overestimated & 45.4\% & 91.0\% & 1.0--2.0 & 12.2\% & 32.9\% \\
    Underestimated & 7.2\% & 47.9\% & 2.0--3.0 & 13.8\% & 23.6\% \\
    Level missed by $\geq1$ & 50.4\% & 76.9\% & $> 3.0$ & 14.6\% & 42.5\% \\
    Level missed by $\geq2$ & 5.0\% & 41.6\% & & & \\
    \bottomrule
  \end{tabular}
  \end{adjustbox}
\end{table}

\paragraph{Some errors are pure arithmetic.} Under CoT prompting, a model's vector and its final score sometimes disagree. For the vector \texttt{AV:N/\allowbreak AC:L/\allowbreak PR:N/\allowbreak UI:N/\allowbreak S:C/\allowbreak C:H/\allowbreak I:H/\allowbreak A:H}, whose correct score is 10.0, both DeepSeek-R1 and LLaMA-3.1-405B answered 9.8. DeepSeek-R1 deferred to a recalled ``official calculator'' value, and LLaMA-3.1-405B invented a numeric weight for Scope. Such errors are removable without changing the model: recomputing $\tilde{s}=f(\vh)$ from the model's own vector eliminates them by construction.

\paragraph{Fine-tuning helps but does not close the gap.} With the same configuration (No.~6 in Supplement~E), fine-tuned LLaMA-3.1-8B reaches 50.00\%, Mistral-7B 47.50\%, Qwen2.5-7B 47.08\%, and Gemma-7B 34.17\% (0-shot prompt; at most four invalid outputs). One further round of training raises Qwen2.5-7B to 56.25\%. Because this configuration was selected on the benchmark (Sect.~\ref{sec:predictors}), 56.25\% is an optimistic upper bound, and even it leaves 44\% of levels wrong.

\begin{answerbox}
\textbf{Answer to RQ1.} Under prompting, at most half of the reports receive the correct level, ten points above always predicting Medium; fine-tuning reaches 56.25\% only with a configuration selected on the benchmark. Errors are directional: every prompting configuration overestimates at least 45\% of reports, and up to 42\% of predictions miss by two or more levels.
\end{answerbox}

%% file: sections/5-rq2.tex
\section{RQ2: Where Do Errors Concentrate?}
\label{sec:rq2}

\begin{figure}[t]
  \centering
  \includegraphics{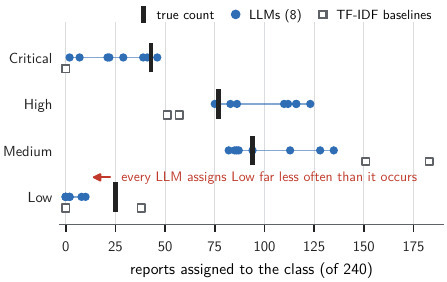}
  \caption{How many reports each predictor assigns to each severity class under few-shot prompting, against the true count (black bar). Counts are derived from per-class precision and recall over valid outputs; the None class (one report) is omitted.}
  \label{fig:classes}
\end{figure}

\paragraph{Severity levels: low severity is almost invisible.} Per-class recall under few-shot prompting (Supplement~G) shows that, of the 25 Low reports, no LLM recognizes more than one (recall $\leq 0.04$), and the single None report is never recognized, while Medium and High are recalled at 0.34--0.68. Fig.~\ref{fig:classes} shows where the missing Low reports go: every LLM assigns the Low class to at most ten reports, against 25 that belong there, while the High class receives 75--123 reports against 77, and seven of eight models fill it beyond its true size. Low reports are thus not misjudged by a small margin but absorbed into higher classes, the class-level image of the overestimation in RQ1. At the top of the scale, models differ in calibration: Critical receives between 2 (Qwen2.5-7B) and 46 (GPT-3.5-turbo) reports against 43. The TF-IDF baselines fail differently, collapsing onto Medium (151--183 reports) and never correctly identifying a Critical report, so the upward drift is specific to the LLMs.

\begin{insightbox}
\textbf{Insight 2: a low score is a claim of absence.} A high score needs one plausible exploit path; a low score needs the absence of worse paths, which a report rarely establishes. Upward drift is thus expected whenever such evidence is missing, not only a calibration fault \op{2}.
\end{insightbox}

\begin{figure}[t]
  \centering
  \includegraphics{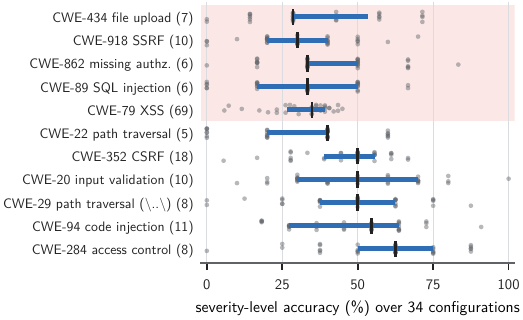}
  \caption{Severity-level accuracy by CWE type (benchmark reports in parentheses). Each dot is one of the 34 configurations; bars span the interquartile range and black ticks mark the median. Shaded: the five hardest types, whose severity depends on context a report rarely states.}
  \label{fig:cwe}
\end{figure}

\paragraph{Vulnerability types: context-dependent classes are hardest.} Level accuracy for the eleven CWE types with at least five benchmark reports, across all 34 configurations (Fig.~\ref{fig:cwe}) has a median ranging from 28.6\% for unrestricted file upload (CWE-434) to 62.5\% for access control (CWE-284). The hardest classes, file upload, SSRF (CWE-918), missing authorization (CWE-862), SQL injection (CWE-89), and XSS (CWE-79), share a property: their severity depends on context that a report rarely states in full, such as what an uploaded file can execute, which internal services a forged request can reach, or in which context a script renders~\cite{fei2025large}. These are the cases in which the model must decide what to assume about missing information, the situation addressed by boundary B3 in Sect.~\ref{sec:rq3}. Sample sizes per class are small (5--18 except XSS with 69), so differences between neighbouring classes should not be over-interpreted; the full table is in Supplement~F.

\begin{answerbox}
\textbf{Answer to RQ2.} Errors are not spread evenly. Low and None reports are almost never recognized and are absorbed into higher classes, and vulnerability types whose severity depends on context the report leaves unstated are the hardest.
\end{answerbox}

%% file: sections/6-rq3.tex
\section{RQ3: Which Semantic Boundaries Break?}
\label{sec:rq3}

\begin{figure*}[t]
  \centering
  \includegraphics[width=0.8\textwidth]{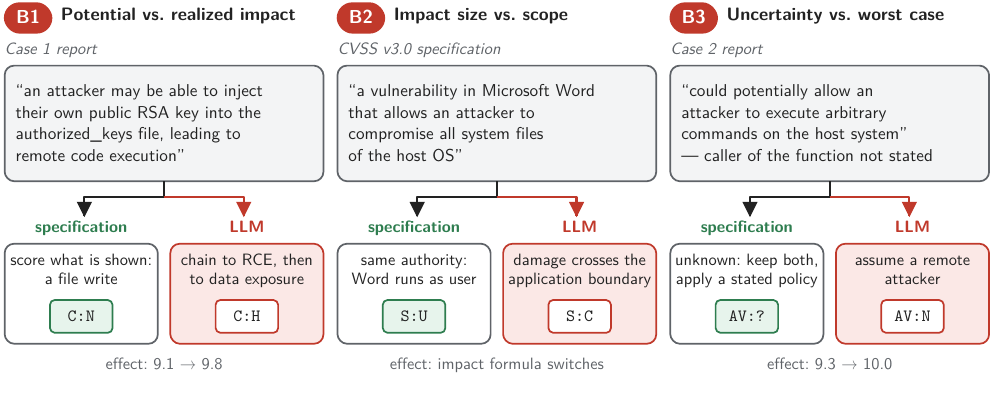}
  \caption{Three semantic boundaries that LLMs violate. Each panel starts from a real sentence (two from analysed reports, one from the CVSS specification) and contrasts the reading the specification requires with the reading the models adopted; the effect line gives the resulting change of the score.}
  \label{fig:boundaries}
\end{figure*}

To explain the patterns of RQ1 and RQ2, we analysed the rationales that models give for their sub-vector choices. We examined four reports in depth: three sampled at random (a path traversal, CWE-22; an OS command injection, CWE-78; a code injection, CWE-94) and the benchmark's single None-severity report. For each, we compared all models' sub-vector rationales with the ground truth, asked follow-up questions where they disagreed, and read the reasoning traces of DeepSeek-R1 (details in Supplement~I). We then tested one recurring error against an example from the CVSS specification itself. The errors fall into three violations of boundaries that the CVSS specification draws but that narrative reports blur (Fig.~\ref{fig:boundaries}). The taxonomy was derived from these cases; its benchmark-wide support is indirect, through the direction of errors (RQ1) and the class and CWE patterns (RQ2), and a direct count over all model rationales is left open (OP5).

\paragraph{B1: potential vs.\ realized impact.} CVSS asks assessors to score the impact a report substantiates under the stated conditions. Models instead escalate along chains the report only suggests. In Case~1, a path traversal allows files to be overwritten, and the impact section notes that an attacker could, for instance, place SSH keys or a webshell. The ground truth is C:N, because no data is disclosed. Every model that produced a usable rating, including DeepSeek-R1, GPT-4o, and LLaMA-3.1-405B, rated confidentiality High; DeepSeek-R1's trace reasons that overwriting enables remote code execution and therefore data exposure. Conditional phrases such as ``could lead to'' are thus read as evidence that the downstream compromise happened.

\paragraph{B2: impact magnitude vs.\ authorization scope.} Scope captures whether an exploit crosses a security authority, not how severe the outcome is. Models treat severity as evidence of a crossing. We tested this with an example from the CVSS v3.0 specification: a vulnerability in Microsoft Word that lets an attacker compromise all system files of the host OS. The specification rates it Scope Unchanged, because the application and the files it accesses fall under the same authority. Every model in our study answered Changed; GPT-4o, for example, argued that the impact ``crosses the application boundary and enters the security boundary of the operating system''. (A current frontier model answered Unchanged, but the example appears verbatim in the specification and may simply be recalled; Sect.~\ref{sec:frontier}.) The same violation occurs on real reports: for Case~3, labelled S:U, DeepSeek-R1 chose S:C (Supplement~I). Since S:C switches the impact equation (Fig.~\ref{fig:cvss}), one misread metric shifts the whole score upwards.

\paragraph{B3: uncertainty vs.\ worst case.} CVSS treats missing information as unknown and asks assessors to score what is supported. When details are missing, models instead assume the worst plausible outcome. They frequently remark that a report is incomplete and then assign a higher severity; GPT-4o states this conservatism explicitly. Information was missing because details lived in unreachable links, the researcher assumed vendor context, or the report was of low quality~\cite{bettenburg2008makes}. This offers an explanation for the near-zero recall of Low and None reports and for the difficulty of context-dependent CWE types (RQ2): a low score requires the model to conclude that an exploit path is \emph{absent}, and models resolve that uncertainty upwards instead.

\paragraph{Where the violations originate.} In the analysed reports, indirect impact descriptions trigger B1, emphatic impact extent with an unclear authorization context triggers B2, and incomplete information triggers B3; asked how the reports would have to change for the vendor's scores to be correct, DeepSeek-R1 proposed exactly these edits. The three differ in kind: B2 is \emph{interpretive} (the report suffices and the specification states the rule), B3 an \emph{information gap} (e.g., whether a library function is reachable remotely is not in the report), and B1 \emph{narrative-induced} (the report juxtaposes demonstrated and speculative impact). All three originate at least partly in the report, which motivates intervening on its format (Sect.~\ref{sec:structure}).

\begin{answerbox}
\textbf{Answer to RQ3.} LLM errors follow three semantic boundary violations: potential impact read as realized (B1), impact magnitude read as an authorization-scope change (B2), and missing information read as the worst case (B3). In the analysed reports, each is triggered by an identifiable feature, namely indirect impact descriptions, unclear scope, and incomplete information, and each pushes the score upwards.
\end{answerbox}

%% file: sections/7-structured-reports.tex
\section{Exploratory Case Study: Structuring the Report}
\label{sec:structure}

The violations of RQ3 arise in the input, not only in the model: the report mixes demonstrated with claimed impact, leaves the security authority implicit, and omits deployment facts. This suggests exploring the report format, rather than the model, as a variable, much as structured behavioural profiles rather than raw code have been given to LLMs to describe malware~\cite{wu2025promal}. Because undisclosed reports are among a vendor's most sensitive data, the vendor may be limited to small local models or to sharing as little as possible (Sect.~\ref{sec:intro}); under such a capability constraint, the input representation is what the platform controls.

\begin{table}[tb]
  \centering
  \caption{Boundary-aligned report structure, as used in the running examples.}
  \label{tab:schema}
  \footnotesize
  \setlength{\tabcolsep}{3pt}
  \begin{tabular}{@{}lp{0.56\columnwidth}ll@{}}
    \toprule
    Field & Content & Informs & Guards \\
    \midrule
    F1 & affected component; whose authority it runs under and governs & S & B2 \\
    F2 & entry point; who controls the input; realistic caller of library code & AV, PR & B3 \\
    F3 & preconditions: configuration, authentication, victim action & AC, PR, UI & B3 \\
    F4a & attacker primitive obtained (e.g., command execution as process user) & C, I, A & B1 \\
    F4b & effect the PoC actually observed & C, I, A & B1 \\
    F5 & consequences the reporter claims, quoted verbatim & not scored & B1 \\
    F6 & per metric, the facts the report does not establish & any & B3 \\
    F7 & behavioural evidence summary, no exploit code & -- & exposure \\
    \bottomrule
  \end{tabular}
\end{table}

\subsection{A Candidate Report Structure}

Table~\ref{tab:schema} gives a first, untuned report structure derived from the three boundaries. Each field corresponds to the sub-vectors it informs and to the boundary it guards. Claimed consequences are kept, quoted verbatim, in a field that is not scored (F5), which addresses B1 without discarding information; impact is split into the primitive the attacker obtains (F4a) and the effect the PoC showed (F4b). The security authority and the attacker's position are explicit (F1, F2), addressing B2 and the most frequent B3 gap; unknowns are listed per metric (F6), so a model can mark a metric as undetermined instead of resolving it upwards; and the proof of concept is replaced by a behavioural summary (F7), since CVSS scores what an attacker can achieve, not how.

\subsection{Running Examples}
\label{sec:running}

We illustrate the structure on four representative reports (Fig.~\ref{fig:running}; inputs and outputs in Supplement~H). RE1--RE3 are Cases~1--3 of Sect.~\ref{sec:rq3}; RE4 is a Huntr report on a missing password check when changing an account's e-mail address. All four also carry an NVD label. A label-blind model instance converted each raw report into the structure of Table~\ref{tab:schema}, quoting claimed consequences verbatim, listing unknowns per metric, and keeping no exploit code. Each raw report and each view was scored by a capability-limited model (Claude Haiku~5.5, no tools) and a frontier model (Claude Opus~5.5, with a Python interpreter and the CVSS specification), with one prompt that allows marking a metric as undetermined. Model capability and tool access are confounded in this comparison, and this is a qualitative case study on four reports, not a benchmark evaluation.

\begin{figure}[t]
  \centering
  \includegraphics{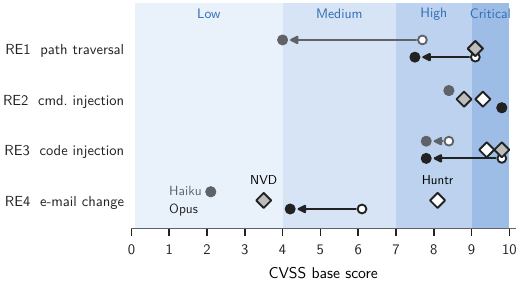}
  \caption{Running examples: each arrow leads from a model's score on the raw report (hollow) to its score on the structured view (filled); upper row Claude Haiku~5.5 without tools, lower row Claude Opus~5.5 with tools. Diamonds mark published labels (white: Huntr; grey: NVD; they coincide for RE1). Vectors are listed in Supplement~H.}
  \label{fig:running}
\end{figure}

\paragraph{Structure dampens B1 but does not remove it.} On the raw RE4 report, the frontier model reasoned that changing the e-mail enables a password reset and hence account takeover, and scored C:H/I:H (6.1). The reporter never claims this chain. On the structured view, whose F5 only quotes the reporter's claim, the same model still mentioned a possible takeover but rated it C:L/I:L (4.2), above the NVD's 3.5. An earlier conversion that stated in F5 that a takeover was not claimed had led to 2.4 (Supplement~H), so part of that effect came from the converter's wording. A stronger model is also better at constructing plausible chains; separating claims from evidence weakens the invitation to do so without removing it.

\paragraph{Structure makes uncertainty explicit but does not settle it.} Once the prompt allows it, both models mark AV and PR as undetermined on every library report (RE1--RE3), with or without structure. The structured views widen this to further metrics, most for the capability-limited model (from 10 to 26 flags over the four reports). The best estimates behind the flags still differ: for RE2 the frontier model applied the user guide's ``reasonable worst case'' for library code~\cite{FIRST2023} and chose AV:N, for RE3 it chose AV:L. Residual disagreement thus concentrates on policy-dependent metrics, which a triage tool can resolve consistently once explicit but no representation settles.

\paragraph{The primitive, not the observed effect.} Separating the primitive (F4a) from the observed effect (F4b) let both models score command execution by what it grants (C:H/I:H/A:H for RE2 and RE3) rather than by a ten-second delay or an empty file. It did not suffice for RE1: the capability-limited model scored the created file (I:L/A:N, 4.0), because F4a grants content control only conditionally and overwriting was not shown.

\begin{insightbox}
\textbf{Insight 3: where the chain is cut is itself a boundary.} Narrative invites chaining too far (B1); restricting models to what the PoC showed makes them stop too early (RE1). The right unit is the capability the attacker holds after the flaw, which CVSS leaves undefined, so every structure must choose a cut point \op{3}.
\end{insightbox}

\paragraph{Less exposure, same information.} No structured view contained exploit code, and the frontier model's raw-to-structured changes (RE1 9.1$\to$7.5, RE3 9.8$\to$7.8, RE4 6.1$\to$4.2) each trace to an explicit field (availability only claimed in F5, the local caller in F2, the takeover absent from F5) rather than to missing exploit detail. Scoring needs the attacker's capability and position, not the payload.

\paragraph{Labels are not neutral either.} Against Huntr labels the structured views look less accurate (Fig.~\ref{fig:running}), but the two published labels disagree on three of the four reports, in both directions: the NVD rated RE2 lower (8.8 vs.\ 9.3), RE3 higher (9.8 vs.\ 9.4), and RE4 far lower (3.5 vs.\ 8.1). For RE4, the Huntr label scores the takeover the reporter does not claim; for RE3, the NVD resolved the unstated caller to the network. Agreement with a single label is therefore an incomplete target.

\begin{insightbox}
\textbf{Insight 4: structure moves the lever from the narrative to the fields.} Because scores follow the fields, the fields now carry the weight the narrative carried. Errors become attributable, but a reporter rewarded by severity also gains a precise place to overstate, and even an exploit-free view reveals component and primitive \op{4}.
\end{insightbox}

%% file: sections/7-discussion.tex
\section{Discussion}
\label{sec:discussion}

\paragraph{Do stronger models make the findings moot?}
\label{sec:frontier}
The evaluated models are not the most capable available today. In our results, however, accuracy does not grow with model capability: the reasoning model DeepSeek-R1 (at most 45.42\%) does not exceed GPT-4o (49.58\%), LLaMA-3.1-405B matches GPT-4o, and examples lift 7B models to 42--44\%, within six points of the best. Because of this, and because the violations we study are rooted in the report and the specification, we did not evaluate the newest and most expensive models; a systematic evaluation of frontier models is planned once funding allows. The running examples, in which a frontier tool-using model computed every score correctly, show how the findings depend on capability; they match the distinction drawn in Sect.~\ref{sec:rq3}.

\begin{insightbox}
\textbf{Insight 5: the origin of an error, not the metric it hits, determines the remedy.} Capability removed the arithmetic errors and, possibly through recall, the interpretive B2. It did not remove the narrative-induced B1, which the frontier model committed on raw RE4 and representation only dampened, and nothing settled B3, which needs facts from outside the report \op{5}.
\end{insightbox}

We therefore regard the three-level evaluation and the boundary taxonomy, rather than any accuracy figure, as the durable contributions: they do not depend on a particular model and matter most where only capability-limited models may be used.

\paragraph{Towards mitigation.} Three measures follow, none of which depends on a particular model. (i) \emph{Separate assessment from computation:} the model outputs only the vector and the CVSS formula computes the score, which removes arithmetic errors (RQ1). (ii) \emph{Structure the input} as in Sect.~\ref{sec:structure}, collected as a submission form or derived at intake. (iii) \emph{Flag rather than score:} when F6 leaves a metric open, the tool should report the alternatives under an explicit platform policy, and route High and Critical candidates whose rationale rests on chained impact, a Scope change, or missing information to a human.

\paragraph{Threats to validity.}
\label{sec:threats}
\emph{Ground truth:} labels are negotiated and partly subjective, and published labels disagree (Sect.~\ref{sec:running}), so some ``errors'' are defensible alternatives; the violations we report, however, contradict explicit definitions in the specification. \emph{Contamination:} the reports are public and may have been seen in training, which would raise accuracy but cannot explain systematic overestimation. \emph{Models and prompts:} newer models may score higher, so absolute accuracy figures may date, whereas the taxonomy, the evaluation framework, and the structure do not depend on them; prompt examples avoid stating CVSS rules and are disjoint from the benchmark. \emph{Scope:} one platform, CVSS~v3.x Base metrics, English reports, and small per-CWE samples; fine-tuned accuracy is an upper bound; the boundary analysis and the running examples are illustrative (OP5), and the structure was filled in by a model, not by reporters.

%% file: sections/8-related-work.tex
\section{Related Work}
\label{sec:related}

\paragraph{Automated severity prediction.} Learning CVSS scores from vulnerability descriptions has a long history, from word embeddings with shallow CNNs~\cite{han2017learning} and multi-target sub-vector prediction~\cite{spanos2018multi} to character-level CNNs~\cite{nakagawa2019character} and distilled transformers~\cite{kai2023vuldistilbert}; all require labelled corpora and short, normalized descriptions. Recent work evaluates LLMs on the same inputs: Marchiori et al.~\cite{marchiori2025can} derived CVSS~v3.1 vectors from CVE descriptions and proposed hybrid LLM--embedding pipelines, and Jafarikhah et al.~\cite{jafarikhah2025description} evaluated six LLM families on more than 31,000 CVEs, observing correlated errors across models and attributing them to vague descriptions. Both report per-metric accuracy. We complement them in two ways: we study the full researcher-written reports that triage actually receives, and we characterize the direction of errors and their grounding in the CVSS specification.

\paragraph{LLMs for vulnerability analysis.} Beyond scoring, LLMs have been evaluated for vulnerability analysis tasks~\cite{yin2024multitask}, code and obfuscated-code analysis~\cite{fang2024large}, penetration testing~\cite{deng2024pentestgpt}, classification of vulnerable agent skills at marketplace scale~\cite{liu2026agentskills}, and vulnerability detection, where several studies find that they reason unreliably about security semantics~\cite{ullah2024llms,ding2025vulnerability,fei2025large}, as learning-based detectors degrade under practical evaluation settings~\cite{wang2025vultegra}. Work that combines LLMs with explicit structure, such as taint specifications for static analysis~\cite{li2024iris} first-order logic models of program behaviour~\cite{yin2025bridging}, or rules extracted from deep anomaly detectors in security applications~\cite{han2024gead}, suggests that reliability improves when the formal part of the reasoning is made explicit. Our findings point in the same direction for severity assessment.

\paragraph{Summary.} Existing evaluations measure how often models are right on short descriptions; how they err on the narrative reports of real triage, and whether those errors are structured enough to be anticipated and mitigated, has remained open.

%% file: sections/8b-open-problems.tex
\section{Open Problems}
\label{sec:open}

\paragraph{OP1: Which accuracy is attainable?} Negotiated labels can inherit the narrative (RE4: Huntr 8.1, NVD 3.5), and no benchmark records how far independent assessors disagree on one report, so a 50\% accuracy cannot be split into model error and irreducible disagreement. Independent vectors with rationales per report would allow evaluation against the set of defensible vectors.

\paragraph{OP2: How can evidence of absence be supplied?} Low scores and most B3 gaps depend on facts outside the report, such as whether a library function is reachable from a network-facing entry point. Program analysis can establish some of them: constraint-guided reachability analysis already decides, for vulnerabilities in dependencies, whether a vulnerable function is reachable or provably unreachable from the application~\cite{feng2026convreach}. Open is how to feed such verdicts into F2, F3, and F6 without anchoring the model on the false positives of static analysers~\cite{li2023sast}.

\paragraph{OP3: What is the unit of attacker capability?} Capability-based security suggests a candidate for the primitive of Insight~3, the authority the attacker holds right after triggering the flaw; whether it can be stated per CWE and checked against F7 is open.

\paragraph{OP4: Is a structured report robust and minimal?} \emph{Manipulation:} can fields be checked against the behavioural evidence, so that an overstated primitive is detectable? \emph{Leakage:} what is the least information that still determines the vector, and how much does it help an adversary rediscover the exploit? This leakage--utility trade-off is unmeasured for vulnerability reports.

\paragraph{OP5: Can boundary violations be detected automatically?} The taxonomy rests on few in-depth cases, and benchmark-wide statistics show the direction of errors but not their cause. Explanation methods for learning-based vulnerability detectors often fail to localize the relevant statements~\cite{cheng2024beyond}, and checking each sub-vector rationale against the report sentence it relies on faces the same faithfulness problem. Such a check, e.g., C:H supported only by a conditional phrase, would nonetheless yield a measured error profile and a run-time guardrail.

%% file: sections/9-conclusion.tex
\section{Conclusion}
\label{sec:conclusion}

We studied how LLMs assess the CVSS severity of real bug bounty reports. Across eight models and four prompting strategies, at most half of the reports received the correct severity level, and fine-tuning raised this only to 56\% with a configuration selected on the benchmark; and the errors were directional: models overestimated far more often than they underestimated and almost never recognized low-severity vulnerabilities. We traced these errors to three semantic boundaries that the CVSS specification draws and narrative reports blur, separating potential from realized impact, impact magnitude from authorization scope, and uncertainty from the worst case.

Because the violations originate partly in the report, its representation is a lever that remains available when the model cannot be upgraded, as is common for sensitive, undisclosed reports. In an exploratory case study, a boundary-aligned, exploit-free structure dampened a narrative-induced error even for a frontier model and made the remaining, policy-dependent uncertainty explicit, although it did not settle it. Evaluating frontier models systematically once funding allows, evaluating such structures at scale with reporters rather than models filling them in, extending them to CVSS~v4.0, and the open problems of Sect.~\ref{sec:open} are the next steps toward triage whose failure modes are known in advance.

%% file: sections/10-data-availability.tex
\section*{Data Availability}
The replication package will be released upon acceptance. The full set includes the identifiers and labels for all 4,281 reports, the benchmark split, and the experiment scripts.

%% file: sections/11-ack.tex
\section*{Acknowledgment}
In line with the IEEE policy on AI-generated content, we disclose that Claude was used to revise the text and to optimize figures from the authors' draft. All content was checked by the authors, who take full responsibility for it. The running examples of Section~\ref{sec:running} use Claude models as study subjects, as described there.